\documentclass[final,5p,times,twocolumn]{elsarticle}

\usepackage{graphicx}
\usepackage{dcolumn}
\usepackage{bm}%
\usepackage{xcolor}

\begin{document}

\begin{frontmatter}


\title{Electronic, magnetic and optical properties of penta-BN$_2$ nanoribbons: a first principles study}

\author[UFRPE]{M. A. L. Dantas}
\author[UFCG1,UFCG2]{N. F. Fraz\~{a}o}
\author[UNB1,UNB2]{David L. Azevedo}
\author[UFRPE,KIT]{Jonas R. F. Lima}
\ead{jonas.lima@ufrpe.br}

\address[UFRPE]{Departamento de F\'{\i}sica, Universidade Federal Rural de Pernambuco, 52171-900, Recife, PE, Brazil}
\address[UFCG1]{Centro de Educa\c{c}\~{a}o e Sa\'{u}de, Universidade Federal de Campina Grande, 581750-000, Cuit\'{e}, PB, Brazil}
\address[UFCG2]{Unidade Acad\^{e}mica de F\'{\i}sica, Universidade Federal de Campina Grande, 58429-900, Campina Grande, PB, Brazil}
\address[UNB1]{Faculdade UnB Planaltina, Universidade de Bras\'{i}lia, 73345-10, Bras\'{i}lia, DF, Brazil}
\address[UNB2]{Instituto de F\'{\i}sica, Universidade de Bras\'{i}lia, 70919-970, Bras\'{i}lia, DF,  Brazil}
\address[KIT]{Institute of Nanotechnology, Karlsruhe Institute of Technology, D-76021 Karlsruhe, Germany}

\date{\today}


\begin{abstract}
The search for new materials is a very intense task in many technological areas. In 2015, a new variant of graphene was proposed, the pentagraphene, which was followed by the propose of a pentagonal boron nitride structure called penta-BN$_2$. Based on these structures, we investigated the electronic, magnetic, and optical properties of penta-BN$_2$ nanoribbons (p-BNNRs) considering four different kinds of edges, carefully closing the valence shells with H atoms to prevent dangling bonds. To achieve this goal, we used first-principles calculations in a density functional theory framework. Our findings showed that the p-BNNRs have a rich magneto-electronic behavior, varying from semiconductor to half-metal. We obtained that they are ferrimagnetic, having an intrinsic magnetism, which allow potential applications in spintronic or spinwaves. From an optical absorption point of view, they mainly absorb at ultraviolet region of the spectrum, especially at UV-B region, which could indicate a potential application as a UV filter.
\end{abstract}

\end{frontmatter}



\section{Introduction}

The discovery of graphene in 2004 \cite{Novoselov}, which was the first two-dimensional (2D) material (one atom thin layer) isolated, gave rise to a whole new field of research and attracted great attention of several scientific fields. Since then, a few dozen of new 2D materials have already been synthesized or exfoliated, and more than a thousand were identified as easily exfoliated \cite{Naguib}. These materials have the most varied properties and have been suggested in all kinds of applications. For instance, graphene is a semimetal with very high carrier mobility and it has the potential to be the host material for the next generation of electronic and optoelectronic devices \cite{RevModPhys.81.109}. It has already been used, for example, in the fabrication of field effect transistors \cite{doi:10.1002/adma.201502544}. Whereas, transition metal dichalcogenides (TMDCs) are semiconductors with good carrier mobility, high current on/off ratio, large optical absorption, and giant photoluminescence, which makes it promising for the fabrication of a variety of devices, such as solar cells, photo-detectors, light-emitting diodes, and phototransistors \cite{CHOI2017116,doi:10.1116/1.4982736,Manzeli2017,kolobov_2018}. On the other hand, the hexagonal boron nitride (h-BN) is an insulator, which is used as dielectrics in electronic devices \cite{doi:10.1002/adfm.201604811}. It also has a high thermal conductivity \cite{doi:10.1021/nn204153h}. We can also mention here the borophene \cite{Mannix1513,Baojie,LI2018282}, silicene \cite{PhysRevLett.108.155501,Lin_2012}, phosphorene \cite{doi:10.1021/acs.nanolett.6b01459,PhysRevMaterials.1.061002,doi:10.1021/acsnano.8b02953} and MXenes \cite{doi:10.1002/adma.201102306,doi:10.1021/nn204153h}.

Beyond these already synthesized materials, there are a plenty of 2D materials with promising properties being proposed theoretically \cite{Xu2018,PONTES2021110210}. Among them, there are materials with a pentagonal structure resembling the Cairo pentagonal tiling. The first one was proposed in 2015 and is called penta-graphene (pG), which is composed entirely by carbon atoms \cite{Zhang2372,KRISHNAN2017465,KRISHNAN2018257,https://doi.org/10.1002/er.4639,SATHISHKUMAR2020123577}. It is a semiconductor, so there is no need to functionalize it in order to open an energy gap, as occurs with graphene. Then, in 2016, it was proposed a hydrogenated pG called penta-graphane \cite{doi:10.1080/14686996.2016.1219970}, which demonstrated greater stability than the pristine pG. A pentagonal boron nitride structure was also proposed \cite{Li}. The most stable one is called penta-BN$_2$, where each pentagon in the structure is composed by two boron and three nitrogen atoms. In contrast to h-BN, the penta-BN$_2$ is metallic, but the influence of a substrate may turn it in a semiconductor. Also, there is the penta-Pt$_2$N$_4$ \cite{C8NR05561K}, which is considered an ideal 2D material for nanoelectronics, since it has high carrier mobility, large direct band gap up to 1.51 eV and also dynamical, thermal, and ambient stability. The list of pentagonal 2D structures also includes penta-AgN$_3$ \cite{doi:10.1063/1.4930086}, penta-SiC$_2$ \cite{C6RA10376F}, penta-CN$_2$ \cite{doi:10.1021/acs.jpcc.5b12510}, TiC$_2$ \cite{ccc} and PdSe$_2$ \cite{doi:10.1021/jacs.7b04865}, the last one being an experimental investigation.

\begin{figure*}[t]
\includegraphics[width=\linewidth]{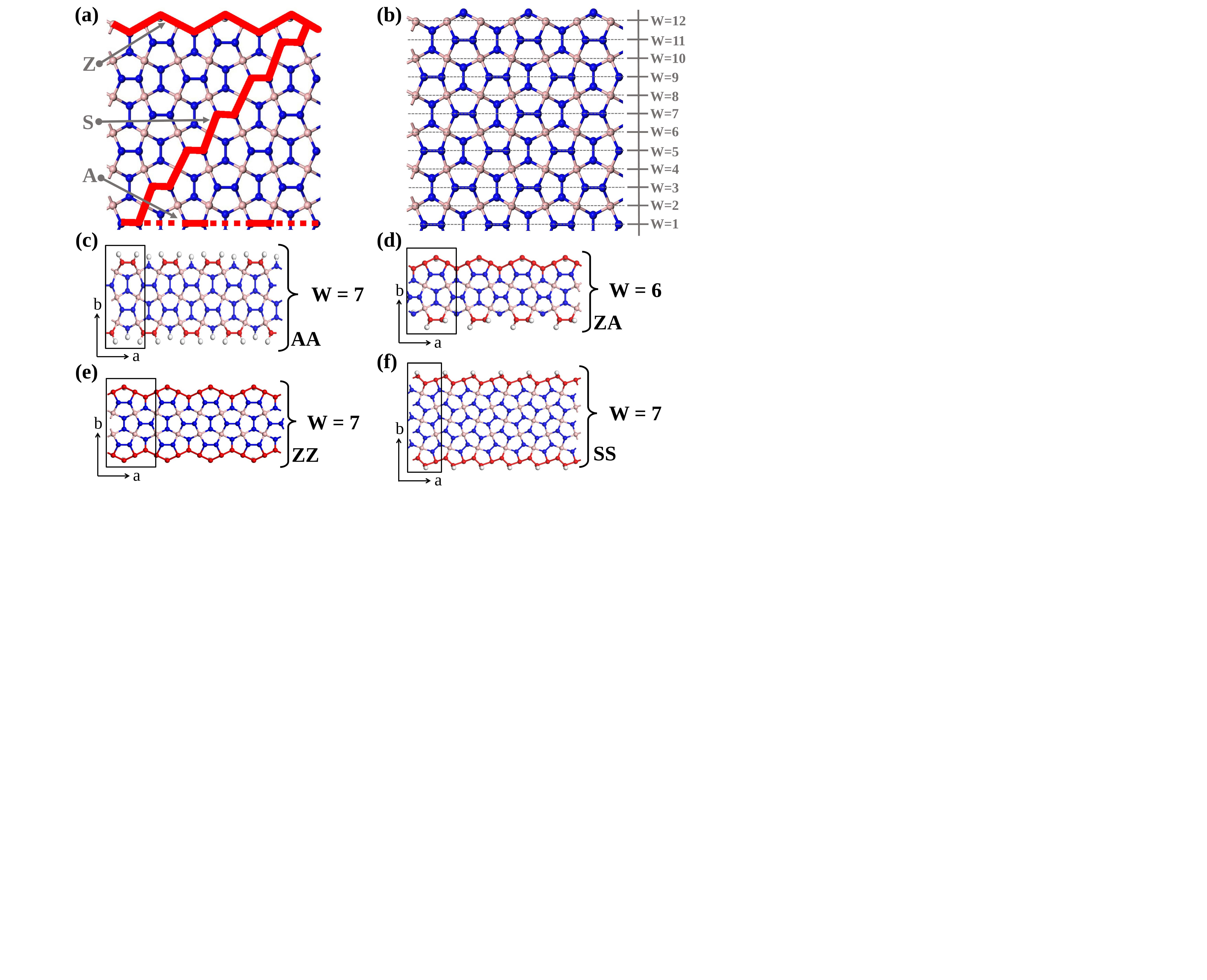}
\caption{\small The penta-BN nanoribbons in its different combinations of edges: (a) penta-BN crystalline structure and its respective cuts, represented by the red segments that delimit them. (b) width of the nanoribbons \textbf{W} is determined as the number of BN strands in the width direction. The results of four typical nanoribbons, with different edge topology, are indicated as (c) AA, (d) ZA, (e) ZZ and (f) SS. The rectangular frames indicate the unit cell used for each nanoribbon and the a-, b-axis are the unit cell vector direction.} \label{fig:01}
\end{figure*}

In the last years, the study of the nanoribbon topology of these 2D materials has attracted great deal of attention \cite{Ersan_2014,doi:10.1080/14786435.2016.1189101,doi:10.1021/acs.nanolett.9b00271,ALESHEIKH2018595,doi:10.1002/admi.201901333,Yamacli2014}. Such study is very important for two main reasons: First, it is one of the arrangements that will effectively be used in the fabrication of nanodevices. Second, the properties of nanoribbons may be different from those of the bulk structure \cite{Lu_2014,doi:10.1021/jp505257g}. For instance, while graphene is a semimetal with zero gap, a graphene nanoribbon with an armchair edge can be metallic or semiconductor depending on the width of the nanoribbon \cite{brey2006electronic,5640645}.

Since the nanoribbon topology for penta-BN$_2$ has not been considered yet, in this paper, we investigate the electronic, magnetic and optical properties of p-BNNRs through first-principles calculations based on the DFT. We obtained that, depending on the edge of the p-BNNR, it can show semiconductor or half-metal behavior. The study of the magnetic properties revealed that the p-BNNR is a ferrimagnetic material, having an intrinsic imbalance between spin up and down. Also, we verified that the optical absorption has peaks in the ultraviolet region of the electromagnetic spectrum, more specifically in the UV-B region, and it is almost transparent in the visible region, revealing that the p-NBNR can potentially be used, for instance, as a protection against UV radiation, which is naturally harmful.  

\section{Method and Structure}

We consider the p-BNNR with four different edges, which can be seen in Fig. \ref{fig:01}. These edges are the same that were considered in the investigation of the magneto-electronic properties of pG nanoribbons \cite{C7CP00029D}. For the first principle calculations of the different p-BNNRs, we used the CASTEP module of Material Studio suite, which is based on the DFT formalism. We replace the core electrons by a Norm-conserving pseudopotential produced by OPIUM. The valence electrons were described as a set of plane-wave basis functions. We choose the well-established Generalized Gradient Approximation (GGA) with the exchange-correlation functional of Perdew-Burke-Ernzerhof (PBE). 

\begin{figure}[]
\includegraphics[width=0.9\linewidth]{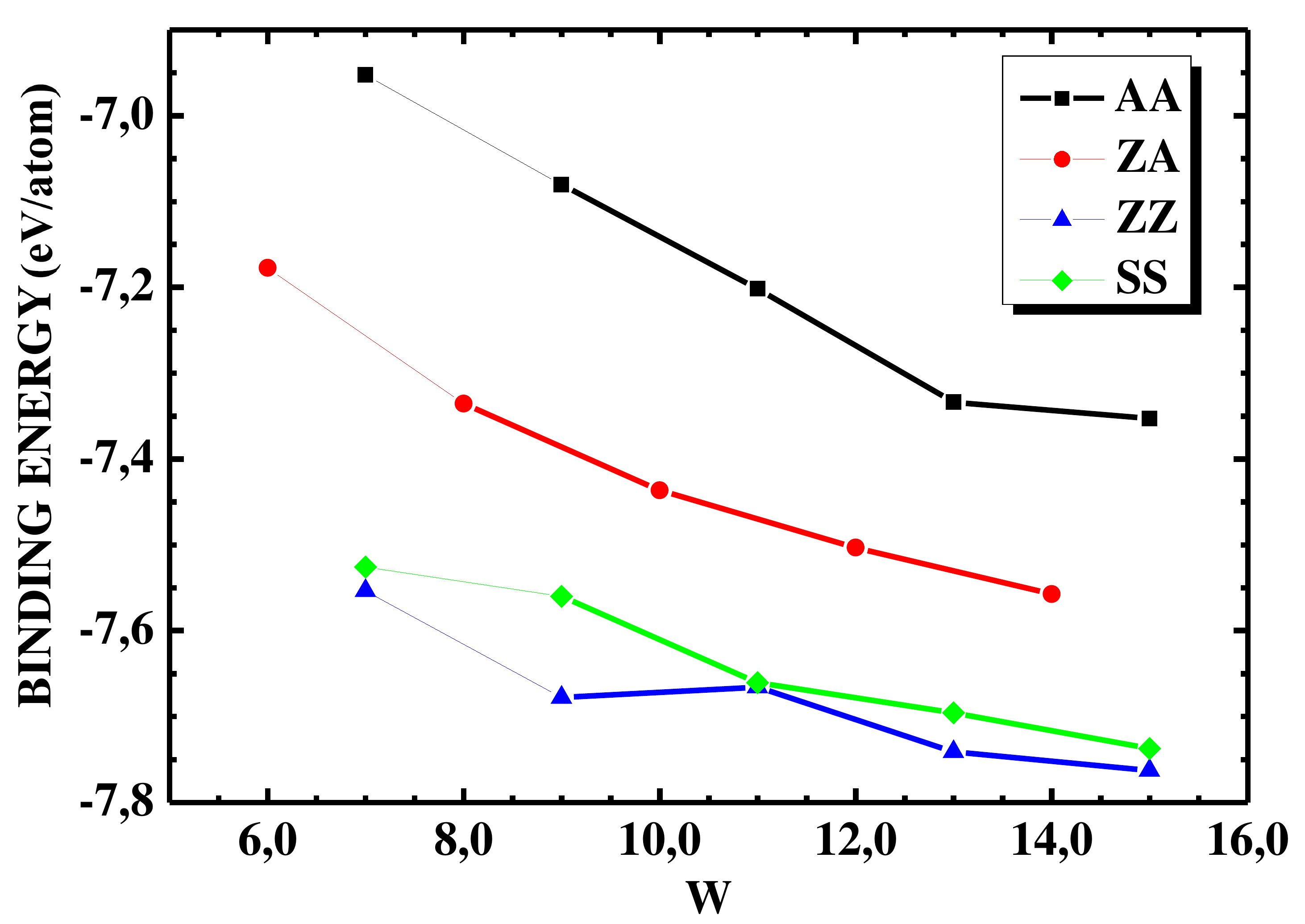}
\caption{\small The binding energy as a function of the width $W$ for the four nanoribbons considered here.} \label{estab}
\end{figure}

On top of the first Brillouin zone was effectuated the integration
by means a k-points sampling using a 7x7x7 Monkhorst-Pack grid, to
ensure that the electronic structure is well converged. To reach the
optimization of the internal atomic positions, the lattice
parameters were released to vary, in order to obtain the total
energy minimization of the unit cells for the p-BNNRs. The convergence parameters for all geometries optimizations were: total energy variation smaller than $2.0\times10^{-5}$ eV, maximum force smaller than 0.05 eV/{\AA}, maximum stress component smaller than 0.1 GPA and maximum
displacement smaller than $2.0\times10^{-2}$~\AA. The tolerance
used for the convergence was of 3 steps, and the optimization method
used was the BFGS minimizer. In the BFGS scheme a starting Hessian is recursively updated. The quality of the base set was kept
fixed, while the volume of the unit cell has undergone changes. For each self-consistent field step, the electronic minimization parameters were total energy/atom convergence tolerance of $1.0\times10^{-5}$ eV  and $0.2174\times10^{-6}$ eV for the electronic eigen-energies. After the optimization, we verified that there is no reconstruction in the edges, only a small deformation in the edge A and Z. From the optimized unit cell with respect to energy for each edge topology, we obtain the binding energy and the band structures. The optical absorption and the complex dielectric function for each structure were calculated, considering a polarized light emitted in a polycrystalline sample.

\section{Results and discussion}

\subsection{Binding energy}

We first evaluated the stability of all p-BNNRs, calculating their binding energy $E_{b}$, defined as  
\begin{equation}
E_{b}= \frac{E_{T}-n_{N}E_{N}-n_{B}E_{B}-n_{H}E_{H}}{n_{N}+n_{B}+n_{H}},
\end{equation}
where $E_{T}$ represents the total energy of the supercell, $E_{N}$, $E_{B}$ and $E_{H}$ are the energies of an atom of N, B and H, respectively, and $n_{N}$, $n_{B}$ and $n_{H}$ are the amounts of each atom inside the supercell, in such a way that $n_N+n_B+n_H$ gives the total quantity of atoms. The binding energy of all nanoribbons for different values of $W$ is presented in Fig. \ref{estab}. One can note that the binding energies range from -7.0 to -7.8 eV, which are well considerable magnitudes, revealing a good stability of these nanoribbons. For the sake of comparison, for pG nanoribbons with widths around the same values considered here, the binding energies range from -7.9 to -8.5 eV, which are close to the values we obtained for the p-BNNRs.

We can also note that the binding energy decreases as the width of the nanoribbon increases, which is a behavior commonly observed in nanoribbons. It means that, as the width of the nanoribbon increases, it becomes more stable. The ZZ nanoribbon has the lowest binding energy, revealing that it is the most stable nanoribbon compared to the other three considered here. It is followed by the SS nanoribbon, which has binding energy very close to the ZZ. This higher stability is a consequence of the predominance of $sp^3$ hybridization in the S and Z edges.

For all performed calculations that follows, we will consider the most stable nanoribbon considered in Fig. \ref{estab} for each edge, which is $W=15$ for the AA, ZZ and SS nanoribbons and $W=14$ for the ZA nanoribbon.

\begin{figure*}[t]
\includegraphics[width=\linewidth]{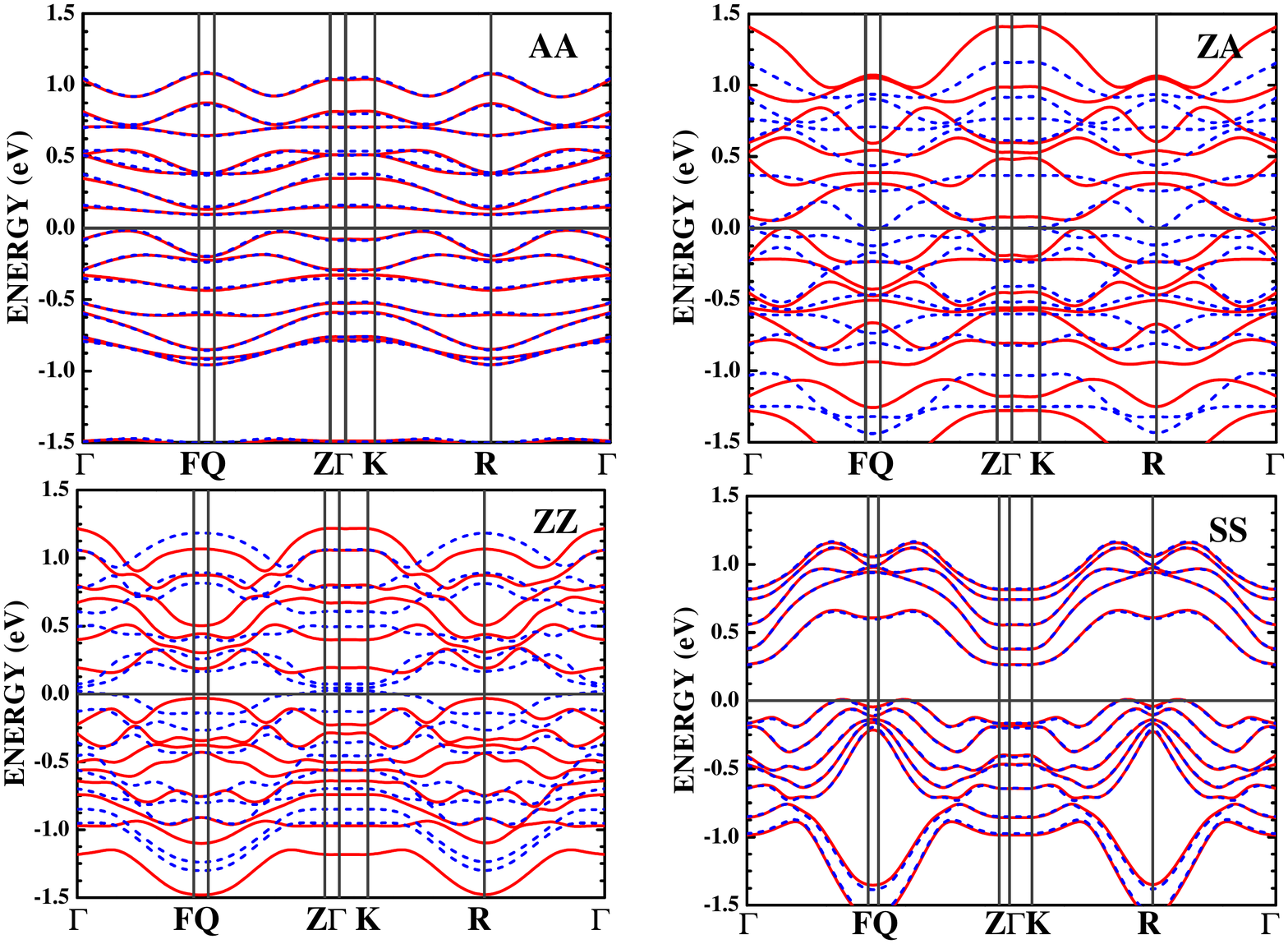}
\caption{\small Kohn-Sham electronic band structure of p-BNNR for each topology considered here, with the red solid lines representing the spin-up state and blue dashed lines the spin-down.}
\label{fig:03}
\end{figure*}


\subsection{Electronic properties}

The electronic structure of Kohn-Sham produces a representation of the electronic eigenenergies, that are periodic functions of the wave vector \(\vec{k}\). In this way, we can describe all the eigenvalues with index \(\vec{k}\) restricted to the first zone of Brillouin (BZ). For the nanoribbons of Fig. \ref{fig:01}, the paths used in the BZ were chosen using straight segments linking a set of high symmetric points. The chosen points are the same for all nanoribbons and are given by: $\Gamma$(0.000, 0.000, 0.000), F(0.500, 0.000, 0.000), Q(0.500, 0.500, 0.000), Z(0.000, 0.500, 0.000), $\Gamma$(0.000, 0.000, 0.000), K(0.000, 0.000, 0.500), R(0.500, 0.000, 0.500), $\Gamma$(0.000, 0.000, 0.000).

The spin-dependent band structure (BS) for low-energy electronic states in the absence of any external electric field for all nanoribbons considered here are shown in Fig. \ref{fig:03}, where the red solid (blue dashed) lines represent the spin up (down). The Fermi level (horizontal black line) was shifted to zero in all the nanoribbons.

It can be observed that the AA and SS nanoribbons are narrow-gap semiconductors, since they present an energy gap smaller than 1 eV. One can note that, for the SS nanoribbon, there is no difference between the band structure for the spin up and down and there is only a negligible difference for the AA nanoribbon. On the other hand, for the ZA and ZZ nanoribbons, the band structures for the spin up and down are not the same. In fact, these nanoribbons are metallic for the spin down and semiconductor for spin up, which means that they present half-metallicity. This reveals the existence of a magnetic characteristic of these structures, since the half-metal materials are ferromagnetic or ferrimagnetic.


\begin{figure*}[t]
\includegraphics[width=\linewidth]{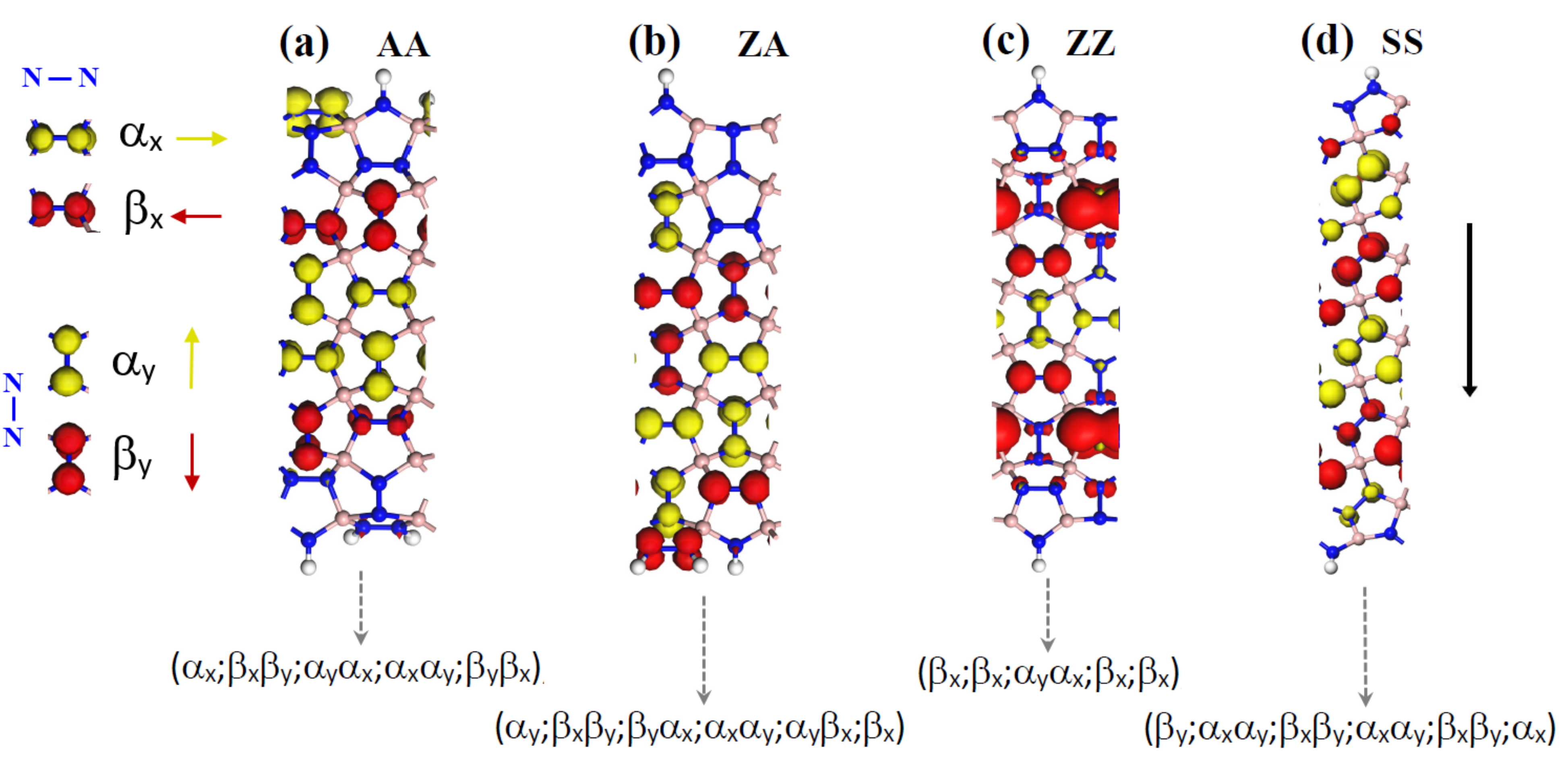}
\caption{\small Theoretical spin-density distribution in four different penta-BN$_2$ nanoribbons identified by the edge shapes. The yellow and red colours denote the $\alpha$- and $\beta$-spin components, respectively. The local magnetic moment is distributed mainly on the specific N-N atoms at horizontal ($\alpha_x$ and $\beta_x$) and vertical ($\alpha_y$ and $\beta_y$).} \label{fig:02}
\end{figure*}

\begin{table*}[!ht] \small
\centering
\begin{tabular}{cccccc}
\hline
\textbf{Parameter}     &  \textbf{Ferrimagnetic}  & \textbf{AA} & \textbf{ZA} & \textbf{ZZ} & \textbf{SS} \\ \hline
2 $*$ Integrated spin density   &  Nonzero & $-0.1997\times 10^{-13}$ & $-0.1013 \times 10^{-12}$ & $-0.9926 \times 10^{-13}$ & $-0.3004\times 10^{-12}$ \\ 
2 $*$ Integrated |spin ~ density |  &   Nonzero, larger magnitude & $8.0309$ & $7.7742$ & $6.0284$ & $6.8518$  \\ \hline
\end{tabular}
\caption{The Integrated Spin Density" and ``Integrated $ \left | spin ~ density \right | $" for all nanoribbons. The comparison of these two values reveals that the nanoribbons are ferrimagnetic.}
\label{table1}
\end{table*}

\subsection{Magnetic properties}

In Fig. \ref{fig:02} is depicted the spin density for the AA, ZA, ZZ, and SS nanoribbons. Analyzing the results, we see that their moments are mostly localized at the N-N atoms, which have a $sp^2$ hybridization. For the spin distribution in the p-BNNRs, four possible spin alignments were obtained: ($\alpha_x;\beta_x\beta_y;\alpha_y\alpha_x;\alpha_x\alpha_y;\beta_y\beta_x$), ($\alpha_y;\beta_x\beta_y;\beta_y\alpha_x;\alpha_x\alpha_y;\alpha_y\beta_x;\beta_x$),
($\beta_x;\beta_x;\alpha_y\alpha_x;\beta_x;\beta_x$), and ($\beta_y;\alpha_x\alpha_y;\beta_x\beta_y;\alpha_x\alpha_y;\beta_x\beta_y;\alpha_x$), where $\alpha_x$, $\alpha_y$, $\beta_x$ and $\beta_y$ denote the different spin orientations and $x$ and $y$ subscript correspond to the direction of N-N bond orientation. Even though we are considering all nanoribbons with valence shell closed by hydrogen atoms, the simulations always converge to a spin-polarized state, suggesting that such nanoribbons have an intrinsic magnetism. However, nanoribbon SS has the spins oppositely-orientated at the structure, so the magnetic moments cancel out completely. In contrast, nanoribbons AA, ZA, and ZZ maintain the magnetic moments different from zero after full convergence, indicating that these three p-BNNRs are systems with interesting magnetic properties. In addition, nanoribbons ZZ exhibits the most localized $\beta_x$ spin density. 

Regarding the magnetic configuration of the AA, ZA, and ZZ nanoribbons, there are ferrimagnetic (FRM) coupling through the p-BNNR width, which can be confirmed comparing the values of the Integrated Spin Density and Integrated $ \left | spin ~ density \right | $ showed in Table 1. Some authors use these values to determine the magnetic moment, but these calculations fail when using ferrimagnetic and anti-ferromagnetic structures \cite{Evans_2014,Goldsby2015}. AA and ZA magnetic states have the same energy in each nanoribbon, implying that the magnetic coupling between the two spin-polarized N-N atoms subsequent is negligible, as happened with the bare single-layer h-BN nanoribbon \cite{doi:10.1021/jp8079827}. As we can see in Fig. \ref{fig:02}, for AA and ZA p-BNNRs, the spin density distributions for the spin up and down are only slightly different, but the magnetic moments in nanoribbons ZZ is preferentially localized in its central part, with $\beta$ spin alignments oriented in $x$-direction. Interestingly, multiple electronic configurations are observed in the p-BNNRs, such as AA (magnetic semiconductor, MS), ZA (half metal, HM), ZZ (half metal, HM), and SS (semiconductor, S), as shown in the spin-polarized BS plot of the p-BNNR structures. Thus, ZZ p-BNNR promises richer magneto-electronic properties than graphene and penta-graphene nanoribbons. It is very promising material for spintonic or spinwaves transport applications.

\begin{figure*}[t]
\includegraphics[width=\linewidth]{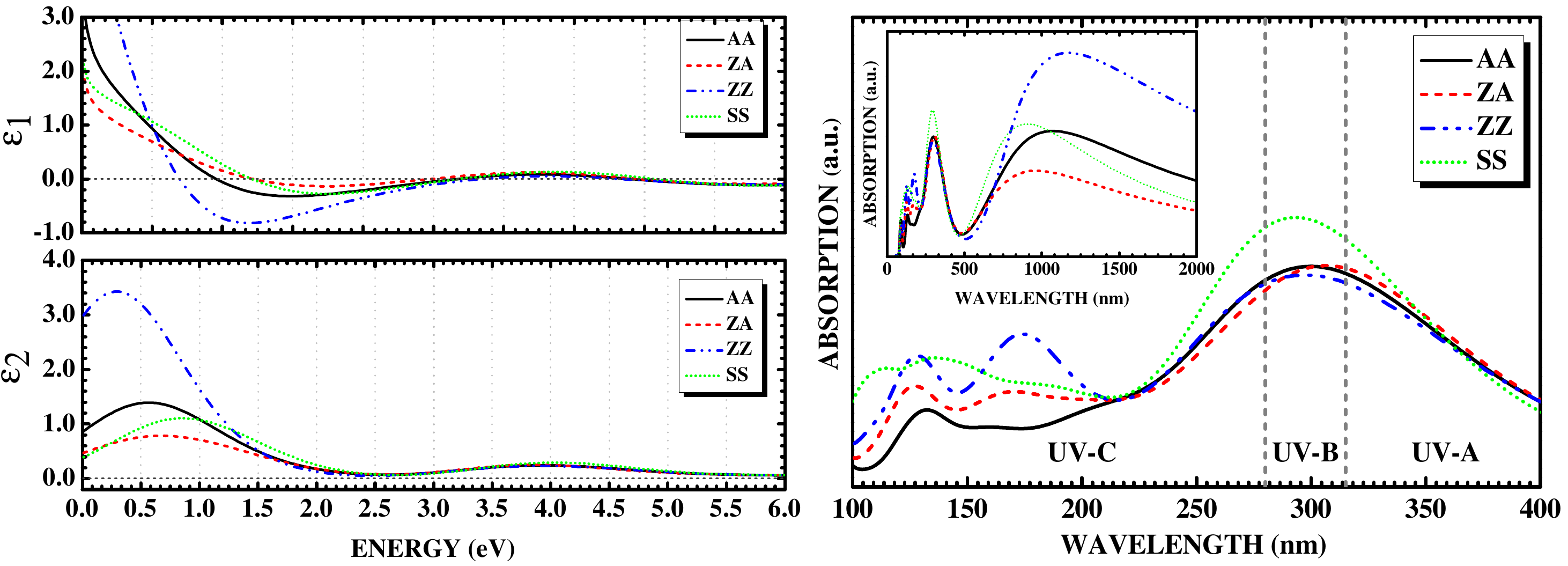}
\caption{\small On the left panel are depicted the real parts ($\varepsilon_{1}$) and the imaginary parts ($\varepsilon_{2}$) of the dielectric function of the four p-BNNR nanostructures. Curves are shown for light incident on a polycrystalline (poly) sample. Optical absorption of the four p-BNNR structures is presented on the right panel, where the vertical dotted lines define the limits of the UV-A, UV-B, and UV-C regions. The inset plot in Absorption graph covers a range of 0 - 2000~nm.} \label{fig:05}
\end{figure*}

In order to understand the magneto-electronic properties observed in these nanoribbons let us first consider the difference in electronegativity in the atoms of the system, which concentrate more electrons around the nitrogen atoms. As was calculated in Ref. \cite{Li}, the Bader charge analysis in the penta-BN$_2$ says that the boron atom has 1.0 electron, while the dinitrogen has 12.1. It means that each dinitrogen in the penta-BN$_2$ has two extra electrons in its vicinity. Our calculations reveal that these extra electrons around two neighbor nitrogen atoms have the same spin orientation. It explains the spin density appearing only in the nitrogen atoms. The difference in the spin distribution for the four edges considered here can be understood looking to the hybridization of the boron atoms. The boron atoms in the bulk of the p-BNNRs have a sp$^3$ hybridization and make four bonds, while the boron atoms in the edges have a sp$^2$ hybridization and make three bonds. In the unit cell of the A-edged p-BNNRs there is no boron atoms in the edge, while the edge S and Z have one and two boron atoms, respectively. As can be seen, the spin density is different from zero only in the dinitrogen bonded to boron atoms from the bulk of the nanoribbon, which means that the spin polarization is a consequence of the bond of the sp$^2$-hybridized dinitrogen with the sp$^3$ boron atoms. The imbalance in the spin density in the ZZ p-BNNRs is a consequence of the presence of the sp$^2$-hybridized boron atoms in the edge. 

It is important to stress that magnetic properties in a BN material were obtained, for instance, in a bare \cite{doi:10.1021/jp8079827,PhysRevB.78.205415,C2RA20306E,doi:10.1021/nl080745j}, Oxygen edge-terminated \cite{doi:10.1021/nl201616h}, doped \cite{ZHANG201524,doi:10.1002/adma.201700695} and fluorinated \cite{Radhakrishnane1700842} h-BN nanoribbon. The p-BNNR is the first boron nitride material revealing intrinsic magnetic properties.

\subsection{Optical properties}

The optical properties can be described through the calculation of the real part ($\varepsilon_{1}$) and the imaginary part ($\varepsilon_{2}$) of the complex dielectric function with respect to energy (eV). Fig. \ref{fig:05} (left panels) shows how these functions behave for the p-BNNR structures.  In order to perform these calculations, we considered a polarized light incident on a polycrystalline sample (POLY). We note that $\varepsilon_{1}$ and $\varepsilon_{2}$, for the POLY direction, are almost independent of the polymorphic form of the material. We measured the $\varepsilon_{0}$ and observe that the maximum absorption occurs in a range from 0.0 to 1.5 eV. Among all studied nanoribbon topologies, the ZZ presented the most intense absorption compared to the others.

The highest value of $\varepsilon_{1}$ is represented by the ZZ structure with $\varepsilon_{0}$ = 10.42. The lowest value obtained was $\varepsilon_{0}$ = 3.03 corresponding to ZA nanoribbon. Also, $\varepsilon_{2}$(w) assumes the higher value in the ZZ nanoribbon, which is 0.29~eV. By the other side, the plasma frequency for all nanoribbons presents a more (less) intense magnitude around 1.21~eV (0.68~eV).

In Fig. \ref{fig:05} (right panel), we consider the optical absorption of the p-BNNRs. As can be seen, for the visible region of the spectrum (in the inset plot), all structures absorb modestly only at 700~nm, with a slightly more intense absorption for the SS nanoribbon. On the other hand, a high peak of absorption occurs in the ultraviolet region for all structures. We highlight that the maximum absorption for all nanoribbons considered here occurs at the UV-B region (at 295.91 nm). However, SS nanoribbon has a more intense absorption in this region compared to the other structures. The peaks in absorption plot are directly related to the imaginary part of the dielectric function, that appear at 0.5~eV. 

Comparing the optical properties of the p-BNNRs with other different hexagonal BN structures, we can see that the transparency in the visible and infrared light and the strong optical absorption in ultraviolet light is observed in both kind of systems \cite{doi:10.1021/jp904246j,PhysRevB.64.035104,Watanabe2004,C7RA00260B,C9CP01038F,Mocci_2019}. However, in hexagonal BN materials, the absorption peak occurs in the UV-C region of the electromagnetic spectrum. Since the UV-C radiation is completely absorbed by the atmosphere, the radiation that causes more damage to skin is the UV-B. It reveals that the p-BNNRs, besides of being promising materials for UV optoelectronic devices, are more suitable to be used, for instance, as a filter for the UV-B radiation.

\section{Conclusions}

In summary, we have investigated the electronic, magnetic and optical properties of penta-BN$_2$ nanoribbons. We performed first-principles calculations based on the density functional theory. For all nanoribbons presented, the binding energy indicates good stability. The magneto-electronic properties showed that the pBNNR can be semiconductor or half-metal, depending on the edge. They are ferrimagnetic materials, with an intrinsic magnetism. The optical properties revealed that all nanoribbons mainly absorb at ultraviolet wavelength, especially in the UV-B region. Potential applications could be in three main areas. First, as magnetic materials. The ZZ pBNNR presents an important ferrimagnetic behavior and the band structure and spin density results indicate that it could be a good candidate for spontaneous magnets. Second, the half-metal behavior of the ZZ p-BNNR turns it a promising candidate for a spin current filter, where the majority carriers would be electrons with spin-down. And finally, all nanoribbons could be used as a UV-B filter, with higher efficiency for the SS case.

{\bf Acknowledgements}:This work was partially supported by Capes, CNPq and Alexander von Humboldt Foundation.





\end{document}